\documentstyle[prl,aps,floats,epsf]{revtex}
\begin{document}
\draft

\twocolumn[\hsize\textwidth\columnwidth\hsize\csname
@twocolumnfalse\endcsname
\title{
Very large dielectric response of thin ferroelectric films with the dead layers
}
\author{A.M. Bratkovsky$^{1}$ and A.P. Levanyuk$^{1,2}$}
\address{$^{1}$Hewlett-Packard Laboratories, 1501 Page Mill Road, Palo Alto,
California 94304\\
$^{2}$Departamento de F\'{i}sica de la Materia Condensada, CIII, Universidad\\
Aut\'{o}noma de Madrid, 28049 Madrid, Spain}
\date{October 31, 2000}
\maketitle

\begin{abstract}

We study the dielectric response of ferroelectric  (FE)
thin films with ``dead" dielectric layer at the interface with electrodes.
The domain structure inevitably forms in the FE film in presence of
the dead layer. As a result, 
the effective dielectric constant of the capacitor $\epsilon_{eff}$ 
increases abruptly when the dead layer is thin and, consequently, 
the pattern of 180-degree domains becomes ``soft".
We compare the exact  results for this problem with 
the description in terms of
a popular ``capacitor" model, which is shown to give
qualitatively incorrect results. We relate the present results
to  fatigue observed in thin ferroelectric films.


\end{abstract}
\vskip 2pc ] % end \twocolumn[...]

\narrowtext
We have shown recently that the dead layer forming at the interface between
ferroelectric (FE) thin film and electrodes has drastic effect on the
electric response of a capacitor\cite{BLprl1}. It has direct bearing on
fatigue observed in FE\ capacitors since in many cases the deterioration of
the switching behavior, like the loss of the coercive force and of the
squareness of hysteresis loop, were attributed to the growth of a ``passive
layer'' at the ferroelectric-electrode interface \cite
{land59,mil90,desu95,lem96}. It is of principal importance that the presence
of a dead layer, no matter how thin in comparison with thickness of the
ferroelectric layer, triggers a formation of the domain structure in FE\
film \cite{BLprl1}. We have shown that when the thickness of the dead layer $%
d$ is not very small, the apparent (net) polarization $P_{a}$ of the
ferroelectric with $180-$degree domain walls follows an approximate relation 
$dP_{a}/dE\propto \epsilon _{g}/d,$ which is in good correspondence with
available experimental data (see e.g. \cite{tag95,gruv96} and references
therein). Importantly, the response of this structure to an external bias
voltage becomes more {\em rigid }when $d$\ increases, i.e. when the dead
layer grows, even in the {\em absence} of pinning by defects.

The implication for real systems is that with the growth of the passive
layer the hysteresis loop very quickly deteriorates and {\em looses its
squareness, }as observed. The approximate $1/d$ dependence of the response
suggests that the effective dielectric constant of the capacitor 
\begin{equation}
\epsilon _{eff}=\frac{4\pi LC}{{\cal A}}  \label{eq:eeff}
\end{equation}
may become {\em very large} when the layer is thin. Here $C$ is the
capacitance of the electroded FE film of area ${\cal A}$, with $L\ $the
separation between electrodes. Indeed, when the dead layer is thin, the
domain width $a$ becomes very large, it grows exponentially with $1/d^{2}$%
\cite{BLprl1}$.$ The response of this domain structure is very soft, and
this should translate into very abrupt increase of the dielectric constant $%
\epsilon _{eff}$ of the capacitor. It is easy to show that in the present
case (180$^{o}$ domains) the linear response is not changed by
electromechanical effect, the change in $\epsilon _{eff}$ only appearing in
quadratic terms in external field, but here we are interested in zero-field
value of $\epsilon _{eff}$ only. We have assumed a quadratic coupling
between the elastic strains and the polarization (as in perovskites). In
this case the linear response of 180$^{0}$ domain structure without pinning
is not affected by intimate contact between the ferroelectric and the dead
layer (excluding mere renormalization of materials constants by homogeneous
stresses). Note that this is invalid in the case of 90$^{o}$ domains \cite
{koukhar00} or a linear coupling between the strain and the polarization\cite
{BLxy}.

Here we address the anomalous behavior of the dielectric constant $\epsilon
_{eff}$ in detail. We also discuss the important issue regarding the
relation of the present results to so-called ``capacitor'' model \cite
{mil90,desu95,tag95}. The ``capacitor'' model (incorrectly) presumes that
the dielectric response of the dielectric layer is not affected by the
presence of the dead layer, so the system looks like capacitances in series.
We show that the effective ``dielectric constant'' of the FE\ layer, $%
\epsilon _{f}$, as found in the ``capacitor'' model\cite{mil90,desu95,tag95}%
, is actually {\em negative}. In spite of this the system remains stable
(stiffness of the domain pattern is positive). 

The reason for this
apparently unusual behavior ($\epsilon _{f}<0$) is that the
``dielectric constant'' of the FE 
layer $\epsilon _{f}$ is the non-local quantity which characterizes the
whole system. This non-local behavior is due to long range Coulomb
interaction, which makes the response rigid even when the FE itself would
have a negative ``dielectric constant'' (cf. Ref.~ \cite{BLbuiltin},
Fig.~1). The ``capacitor'' model neglects this essentially nonlocal behavior
and, therefore, is hardly applicable to the problem of dielectric response
of thin ferroelectric films, and to the problem of fatigue for that matter.
The present considerations remain valid until the period of the domain
structure remains much smaller that the {\em lateral size} of the film (or
the grain size). Those will set some cutoff for the effects considered
below.

We shall find the response of a ferroelectric film under a bias voltage $U$
with thickness $l$ separated from the top and bottom electrodes by passive
layers with thickness $d/2$ (Fig.~1, inset) with the use of the
thermodynamic potential\cite{BLprl1}: 
\begin{eqnarray}
\tilde{F} &=&F_{{\rm s}}+\tilde{F}_{es},  \label{eq:ftil} \\
\tilde{F}_{es} &=&\int dVE^{2}/(8\pi )-\sum_{a}^{\text{electrodes}%
}e_{a}\varphi _{a}.  \label{eq:fes0}
\end{eqnarray}
Here $F_{s}$ is the surface energy of the domain walls, $\tilde{F}_{es}$ is
the electrostatic energy, $\vec{E}$ is the electric field, while $e_{a}$ ($%
\varphi _{a})$ is the charge (potential) on the electrode $a.$ The last term
in Eq.~(\ref{eq:fes0}) accounts for the work of the external voltage
source(s). Note that the results would not change qualitatively is there
were only one dead layer, since the accompanying depolarizing field would
have the same effect.

We are interested in a case when the ferroelectric has the spontaneous
polarization $\vec{P}_{s}\parallel z$ (Fig.~1, inset), i.e. we are
considering the case of $180-$degree domain walls. The uniaxial
ferroelectric film has the dielectric constant $\epsilon _{c}$ ($\epsilon
_{a})$ in the $z$-direction (in the $xy$-plane), and the dielectric constant
of the passive layer is $\epsilon _{g}.$\ We select the $x$-axis
perpendicular to the domain walls. The potential $\varphi ,$ which is
related to the electric field in usual way, $\vec{E}=-\nabla \varphi $,
satisfies the following equations of electrostatics in the ferroelectric and
the passive layer\cite{BLprl1}, 
\begin{equation}
\epsilon _{a}\frac{\partial ^{2}\varphi _{f}}{\partial x^{2}}+\epsilon _{c}%
\frac{\partial ^{2}\varphi _{f}}{\partial z^{2}}=0,\quad \frac{\partial
^{2}\varphi _{g}}{\partial x^{2}}+\frac{\partial ^{2}\varphi _{g}}{\partial
z^{2}}=0,  \label{eq:figap}
\end{equation}
with the boundary conditions $\varphi =-(+)U/2,\quad z=+(-)(l+d)/2$, and 
\begin{equation}
\epsilon _{c}\frac{\partial \varphi _{f}}{\partial z}-\epsilon _{g}\frac{%
\partial \varphi _{g}}{\partial z}=4\pi \sigma (x),\quad \varphi
_{f}=\varphi _{g},\text{ \ at }z=l/2,  \label{eq:BC}
\end{equation}
where the subscript $f$ ($g)$ denotes the ferroelectric (passive layer, or a
vacuum gap). Here $\sigma $ is the density of the bound charge due to
spontaneous polarization at the ferroelectric-passive layer interface, $%
\sigma =P_{ns}=\pm P_{s}$, depending on the normal direction of the
polarization at the interface, alternating from domain to domain, Fig.~1
(inset). Thus, we assume that the absolute value of the {\em spontaneous}
part of the polarization $P_{s}$ is constant in all domains and only its
direction is alternating from domain to domain. We have also assumed a usual
separation of linear and spontaneous polarization, so that the displacement
vector is $D_{i}=\epsilon _{ik}E_{k}+4\pi P_{si}$, where $i,k=x,y,z$, and
the dielectric response $\epsilon _{ik}$ in uniaxial, Fig.~1(inset). Within
this approximation we obtain 
\begin{equation}
\tilde{F}_{es}=\frac{1}{2}\int_{\text{FE}}d{\cal A}\sigma \varphi -{\frac{1}{%
2}}\sum_{a}^{\text{electrodes}}e_{a}\varphi _{a},  \label{eq:Fesmain}
\end{equation}
where $\sigma =\pm P_{s}$ is the density of the {\em bound charge} given by 
{\em only} the spontaneous polarization at the interfaces between the FE and
the dead layer, and the integration goes over these interfaces.
The periodic pattern consists of $c-$domains with widths 
$a_{1}$ and $a_{2}$, and the period $T=a_{1}+a_{2}$. The solution of the
equations (\ref{eq:figap}) is then readily found by Fourier transformation 
\begin{eqnarray}
\sigma (x) &=&\sum_{k}\sigma _{k}e^{ikx},\quad \varphi _{\alpha
}(x,z)=\sum_{k}\varphi _{k\alpha }(z)e^{ikx}, \\
\sigma _{k} &=&\frac{2iP_{s}}{kT}\left[ 1-\exp \left( ika_{1}\right) \right]
,\ \hspace{0.2in}k\neq 0,  \label{eq:sk} \\
\sigma _{k=0} &\equiv &\sigma _{0}=P_{s}\frac{a_{1}-a_{2}}{a_{1}+a_{2}}%
\equiv P_{s}\delta ,  \label{eq:s0}
\end{eqnarray}
where $k=2\pi n/T=\pi n/a,$ $n=0,\pm 1,\ldots $, $a\equiv (a_{1}+a_{2})/2$
and the index $\alpha =f,$ $g$ marks the quantities for the FE layer and the
dead layer. Note that by definition $P_{a}\equiv \sigma _{0}$ is the net
spontaneous polarization of the FE layer. We obtain 
\begin{eqnarray}
\tilde{F}_{es} &=&\tilde{F}_{h}+\tilde{F}_{inh},  \label{eq:fes} \\
\frac{\tilde{F}_{h}}{{\cal A}} &=&\frac{4\pi \sigma _{0}d-\epsilon _{g}U}{%
\epsilon _{g}l+\epsilon _{c}d}\frac{l\sigma _{0}}{2}+\frac{1}{2}\sigma
_{0}^{el}U,  \label{eq:fh} \\
\frac{\tilde{F}_{inh}}{{\cal A}} &=&\sum_{k\neq 0}\frac{4\pi \left| \sigma
_{k}\right| ^{2}}{kD_{k}},  \label{eq:finh}
\end{eqnarray}
where 
\begin{equation}
D_{k}=\sqrt{\epsilon _{a}\epsilon _{c}}\coth \left[ \left( \frac{\epsilon
_{a}}{\epsilon _{c}}\right) ^{1/2}\frac{kl}{2}\right] +\epsilon _{g}\coth
\left( \frac{kd}{2}\right) ,  \label{eq:dk}
\end{equation}
and 
\begin{equation}
\sigma _{0}^{el}=-\frac{\epsilon _{g}}{4\pi }\frac{4\pi l\sigma
_{0}+\epsilon _{c}U}{\epsilon _{g}l+\epsilon _{c}d}  \label{eq:sel}
\end{equation}
is the net charge density induced on the electrode at $z=L/2$, with $\cal{A}$
the area of the film. One can
calculate the apparent dielectric constant of the whole capacitor from 
\begin{equation}
\epsilon _{eff}=4\pi L\sigma _{0}^{el}/U.
\end{equation}
The term $\tilde{F}_{h\text{ }}$in Eq.~(\ref{eq:fes}) is due to the net
polarization of the FE film induced by the bias voltage $U$, with the first
term in $\tilde{F}_{h}$ (\ref{eq:fh}) corresponding to the term $k=0$,
singled out in the stray field energy $\tilde{F}_{inh}$ $($\ref{eq:finh}$).$
With the use of $\sigma _{k}$ from (\ref{eq:sk}) we obtain 
\begin{eqnarray}
\frac{\tilde{F}_{inh}}{P_{s}^{2}{\cal A}} &=&\frac{32a}{\pi ^{2}}%
\sum_{n=0}^{\infty }\frac{1}{\left( 2n+1\right) ^{3}D_{2n+1}}  \nonumber \\
&&+\frac{16a}{\pi ^{2}}\sum_{n=1}^{\infty }\frac{\left( -1\right) ^{n}}{n^{3}%
}\frac{1-\cos \pi n\delta }{D_{n}},  \label{eq:finh1}
\end{eqnarray}
where $D_{n}=D_{k_{n}},$ $k_{n}=\pi n/a.$

The total free energy of the domain pattern per unit area is 
\begin{equation}
\frac{\tilde{F}}{{\cal A}}=\frac{\gamma l}{a}+\frac{\tilde{F}_{h}+\tilde{F}%
_{inh}}{{\cal A}},  \label{eq:ftot}
\end{equation}
where $\gamma =P_{s}^{2}\Delta $ is the surface energy of the domain wall,
with $\Delta $ the characteristic microscopic length\cite{BLprl1}. This free
energy allows one to determine the equilibrium properties and response of
the domain pattern to external field.

We can determine a linear response of the system to bias voltage from the
total energy (\ref{eq:ftot}), which, for small bias $U,$ can be expanded up
to terms quadratic in $U\;$\ and $\sigma _{0},$ 
\begin{equation}
\frac{\tilde{F}}{{\cal A}}=\frac{\tilde{F}_{stray}(a)}{{\cal A}}+\frac{1}{2}%
S\sigma _{0}^{2}-RU\sigma _{0},  \label{eq:ftot2}
\end{equation}
where $\sigma _{0}\equiv P_{s}\delta $, $\tilde F _{stray}(a)$ is the usual
stray energy, $\tilde F_{stray}(a) =\tilde F_{inh}(a; \sigma _{0}=\delta =0)$,
given by the first term in Eq.~(\ref{eq:finh1}). This expression results
from expanding the Eqs.(\ref{eq:finh}),(\ref{eq:finh1}) in powers of $U$ and 
$\delta $ (note that $\sigma _{0}\equiv P_{s}\delta ).$ Here $S$ is the
stiffness of the domain structure with respect to external bias voltage $U$
[we have omitted the constant term $\propto U^{2}$ in Eq.~(\ref{eq:ftot2})].
We have 
\begin{eqnarray}
R &=&\frac{\epsilon _{g}l}{\epsilon _{g}l+\epsilon _{c}d},  \label{eq:lres}
\\
S &=&S_{h}+S_{inh,}  \label{eq:sgen} \\
S_{h} &=&\frac{4\pi dl}{\epsilon _{g}l+\epsilon _{c}d},  \label{eq:sh}
\end{eqnarray}
where $S_{inh}$ is the (negative) contribution of the inhomogeneous (stray)
energy, corresponding to the second term in (\ref{eq:finh1}), 
\begin{equation}
S_{inh}=16a\sum_{n=1}^{\infty }\frac{\left( -1\right) ^{n}}{nD_{n}}.
\label{eq:sinh}
\end{equation}
We can now consider the following limiting cases:

{\em Thick dead layer }($a\ll d).-$ There we can replace both $\coth $ in (%
\ref{eq:dk}) by unity, with the result 
\begin{equation}
\frac{\tilde{F}(\delta =0)}{P_{s}^{2}{\cal A}}=\frac{\Delta l}{a}+\frac{%
28\zeta (3)a}{\pi ^{2}\tilde{\epsilon}},  \label{eq:fkit}
\end{equation}
with $\tilde{\epsilon}=\sqrt{\epsilon _{a}\epsilon _{c}}+\epsilon _{g}.$ The
domain structure has a minimum energy for the equilibrium (Kittel) domain
width 
\begin{equation}
a=a_{K}\equiv \left( \frac{\pi ^{2}\tilde{\epsilon}}{28\zeta \left( 3\right) 
}\Delta l\right) ^{1/2} \propto \sqrt{l},
\label{eq:ak}
\end{equation}
so that $a \sim \sqrt{\Delta l} \ll l$, a usual relation.
Then inhomogeneous contribution to stiffness $S_{inh}\approx $ $-\frac{16\ln
2}{\tilde{\epsilon}}a_{K},$ Eq.(\ref{eq:sinh}), and we obtain from Eqs.(\ref
{eq:sgen}),(\ref{eq:sh}) 
\begin{equation}
S=\frac{4\pi dl}{\epsilon _{g}l+\epsilon _{c}d}-\frac{16\ln 2}{\tilde{%
\epsilon}}a_{K}.  \label{eq:SK}
\end{equation}
When $d\gtrsim a_{K},$ one can neglect the second (stray) term in this
expression for stiffness, thus recovering our approximate expression for the
net spontaneous polarization $P_{a}\equiv \sigma _{0}$ of the ferroelectric
film from (\ref{eq:ftot2}) \cite{BLprl1} 
\begin{equation}
\frac{P_{a}}{U}=\frac{R}{S}\approx \frac{\epsilon _{g}}{4\pi d}.
\label{eq:slope}
\end{equation}
This approximation breaks
down for thinner films but it is obvious that the response becomes {\em %
softer} for {\em thinner} dead layers. The breakdown of this approximate
behavior has also been noticed by Kopal {\it et al.} for similar model\cite
{kopal99}, but they have not analyzed what actually happens to the response
at very small thicknesses $d$ of the dead layer. Note that
$S_{inh}<0,$ Eq. (\ref{eq:sinh}), and, consequently, the following
inequality always holds:  $P_{a}/U > \epsilon _{g}/4\pi d$.

{\em Thin dead layer} ($a\gg d).-$In this case the replacement of $\coth $
by unity is not allowed. For $a\lesssim l$ and $\sqrt{\epsilon _{a}\epsilon
_{c}}=\epsilon _{g}$ we have $D_{n}=\epsilon _{g}\left( 1+\coth kd/2\right) $
and 
\begin{eqnarray}
\frac{\tilde{F}_{inh}}{P_{s}^{2}{\cal A}} &=&\frac{16a}{\pi ^{2}\epsilon _{g}%
}\left[ \frac{7}{8}\zeta \left( 3\right) -Li_{3}\left( e^{-b}\right) +\frac{1%
}{8}Li_{3}\left( e^{-2b}\right) \right] ,  \label{eq: finhas} \\
S_{inh} &=&-\frac{8a}{\epsilon _{g}}\ln \frac{2}{1+e^{-b}},
\label{eq:sinhas}
\end{eqnarray}
where $b=\pi d/a,$ $Li_{n}(z)=\sum_{k=1}^{\infty }z^{k}/k^{n}$ \cite{Li}. By
using the asymptotic expansion of $Li_{n}(z)$ we obtain for $d\ll a$ our
earlier result\cite{BLprl1} 
\begin{equation}
a=\frac{\pi d}{2e^{1/2}}\exp \left( \frac{\epsilon _{g}\Delta l}{4d^{2}}%
\right) ,  \label{eq:aexp}
\end{equation}
which corresponds to very abrupt increase of the domain width to values $%
a\gg a_{K},$ with $a_{K}$ the Kittel width (\ref{eq:ak}), when the dead
layer is very thin, Fig. 1. This approximation, however, is not sufficient
to make an estimate of the stiffness for very thin dead layers. Indeed,
formally the stiffness there becomes small and negative, $=4\pi d\left[
1/\left( \epsilon _{g}+\epsilon _{c}d/l\right) -1/\epsilon _{g}\right]<0 $.
This fact simply indicates that for thin dead layers the stiffness
diminishes, $S\rightarrow 0$ and has to be calculated more
accurately. The exact 
results, illustrating the abrupt softening of the dielectric response for
small $d,$ are shown in Fig.~1.
\begin{figure}[t]
\epsfxsize=3.4in
\epsffile{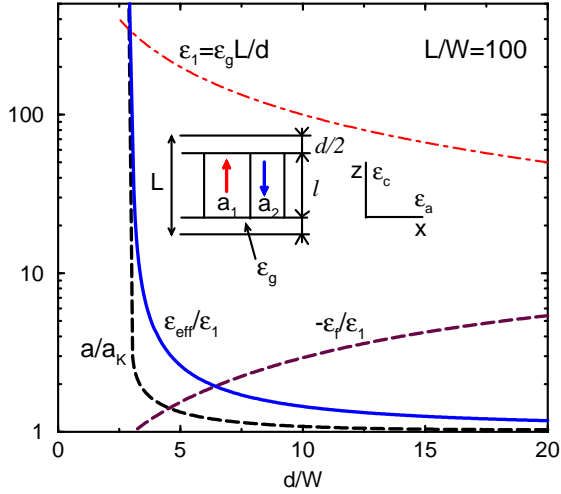}
\caption{ The dielectric constants of the ferroelectric with the dead layers
of thickness $d/2$, as defined in the text. $W$ is the domain wall thickness.
Note an abrupt increase of the effective dielectric constant of the
capacitor $\protect\epsilon _{eff}$ in comparison with $\protect\epsilon %
_{1}=\protect\epsilon _{g}L/d$, and that the ``dielectric constant'' of the
FE layer is {\em negative}, $\protect\epsilon_f <0$. The relation $\protect%
\epsilon _{eff}\approx \protect\epsilon _{1}$, which follows from the
``capacitor'' model, is clearly violated when the dead layer is thin and the
domain width $a$ is large ($a\gg a_K$, the Kittel width, and $\epsilon
_{eff}$ becomes $\gg \epsilon _{1}$). Inset shows 
schematics of the electroded ferroelectric film (capacitor) with the dead
layers.}
\label{fig:fig1}
\end{figure}

One may wish to interpret the approximate result for slope of the net
polarization $P_{a}\propto 1/d,$ given by Eq.(\ref{eq:slope}), in terms of
the ``capacitors in series'' model \cite{mil90,desu95,tag95}. Indeed,
a similar 
result follows if one were to assume that the capacitance of the dead layer, 
$\propto 1/d,$ dominates at small thicknesses of the layer $d.$

An opinion has even been voiced that the effective dielectric constant of
the FE layer is infinite ($\epsilon_f=\infty$?) since the domain walls in our
model are not pinned \cite{tagef00}. However, such an interpretation would
be incorrect since $\epsilon _{f}$, as found in the ``capacitor'' model, is
not infinite, but finite and actually {\em negative}. To establish this, one
has to find the voltage drops across the dead layer and FE film. The
homogeneous part of the electric fields inside the FE (dead) layer $E_{f}$ $%
\left( E_{d}\right) $ and the corresponding voltage drops are found from the
standard equations 
\begin{eqnarray}
E_{d}d+E_{f}l &=&U,  \nonumber \\
\epsilon _{g}E_{d} &=&\epsilon _{c}E_{f}+4\pi P_{a},  \label{eq:maxw}
\end{eqnarray}
where from one can, for instance, easily recover the expression (\ref{eq:sel}%
) for the net charge density on electrodes. One obtains the ``capacitor''
model by assuming that the interface between the FE film and the dead layer
is equipotential and can be viewed as the metallic film separating two
areas. Simple calculation gives 
\begin{eqnarray}
\epsilon _{f} &\equiv &\frac{4\pi l\sigma _{0}^{el}}{U_{f}}=\epsilon _{c}%
\frac{1+4\pi P_{a}l/\left( \epsilon _{c}U\right) }{1-4\pi P_{a}d/\left(
\epsilon _{g}U\right) }<0~(!),  \label{eq:ef} \\
\epsilon _{eff} &\equiv &\frac{4\pi L\sigma _{0}^{el}}{U}=\frac{L\epsilon
_{g}\epsilon _{c}}{\epsilon _{c}d+\epsilon _{g}l}\left[ 1+4\pi P_{a}l/\left(
\epsilon _{c}U\right) \right] ,  \label{eq:etot}
\end{eqnarray}
where $U_{f}\equiv E_{f}l.$ Since always $P_{a}/U>\epsilon _{g}/4\pi d,$
Eq.~(\ref{eq:slope}), we obtain $\epsilon _{f}<0,$ Fig.~1. In spite of
formally negative ``dielectric constant'' of the ferroelectric layer $%
\epsilon _{f},$ Eq.(\ref{eq:ef}), which is an artifact of the ``capacitor''
model \cite{mil90,desu95,tag95,tagef00}, the system remains stable (stiffness of
the domain pattern is positive). Indeed, shifting the domain walls would
create a net electric field in the capacitor, see Eq.~(\ref{eq:fh})
and second term in Eq.~(\ref{eq:ftot2}), and its energy is the source of finite
stiffness of 
the domain pattern, even when the walls are {\em not pinned}, as is
very well known since 1960~\cite{enz60}. 

The reason for this apparently unusual
behavior ($\epsilon _{f}<0$) is that the ``dielectric constant'' of the
FE layer $\epsilon _{f}$ 
in (\ref{eq:ef}) is the non-local quantity which characterizes the whole
system: it depends on the properties of the dead layer. This non-local
behavior is due to long range Coulomb interaction, which makes the response
rigid even when the FE itself would have a negative ``dielectric constant''
(for analogous situation in FE films with depletion charge see Ref.~\cite
{BLbuiltin}, Fig.~1). Thus, the ``capacitor'' model actually operates with
obscure quantities without much physical meaning.

We illustrate the difference between the exact results for $\epsilon _{eff}$
and that following from the ``capacitor'' model $\epsilon _{eff}\approx
\epsilon _{1}\equiv \epsilon _{g}L/d$ in the assumption $\epsilon
_{f}=\infty $ \cite{tagef00} in Fig.~1. The softness of the domain
structure, characterized by $P_{a}/U,$ increases very abruptly when the dead
layer is thin and the width of the domains is $a\gg a_{K}$ [Kittel
width (\ref{eq:ak})].
According to (\ref{eq:etot}), $\epsilon _{eff}$ increases abruptly and
becomes $\gg \epsilon _{1}$ in this region, Fig.~1, in stark deviation from
the prediction of the ``capacitor'' model $\epsilon _{eff}\approx \epsilon
_{1}$\cite{tagef00}. For thick ``dead layers'' the effective dielectric
constant $\epsilon _{eff}$ becomes comparable to $\epsilon
_{1}$, as we showed earlier\cite{BLprl1}.

We think that the above clearly demonstrates a danger of applying a naive
electric circuit analysis to FE systems where the addition of one ``circuit
element'' (dead layer) radically changes the electric response of the other,
FE layer, by introducing a domain structure. It is not surprising,
therefore, that such an approach cannot explain a fatigue observed in FE
films (see e.g. \cite{tag95,gruv96} and references therein). Certainly,
there might be various reasons for the fatigue in FE capacitors. We simply
observe that the growth of the dead layer at the interface with electrode
makes the dielectric response of the film {\em rigid} even when the domain
walls are {\em not pinned}. The present mechanism gives a correct order of
magnitude for the tilt of the hysteresis loops\cite{BLprl1}, therefore the
growth of passive layer might be the main source of fatigue. In the limiting
case of thin dead layers the period of the ferroelectric domains in the
absence of pinning becomes much larger than the standard Kittel width. As a
consequence, the domain pattern becomes very ``soft'' in the absence of the
pinning of the domain walls, and its contribution to the dielectric response
becomes very large, since the domains with polarization parallel to the
external field can easily grow at the expense of the domains with the
opposite polarization.

\end{document}